\documentclass[preprint,aps,floats]{revtex4}
\input epsf.tex
\def\DESepsf(#1 width #2){\epsfxsize=#2 \epsfbox{#1}}

\def\be {\begin{equation}}
\def\ee {\end{equation}}
\def\ba{\begin{array}}
\def\ea{\end{array}}

\def\ra{\rightarrow}

\def\l {\lambda}

\def\betak {B\rightarrow \eta K}
\def\betak0 {B^{0}\rightarrow \eta' K^{0}}

\def\betapkstr0 {B^{0}\rightarrow \eta' K^{*0}}
\def\bpetak0 {B^{0}\rightarrow \eta' K^{0}}

\def\msnus {m_{\tilde\nu_{iL}}^2}
\def\msells {m_{\tilde e_{iL}}^2}

\def\lappeq{\mathrel{\rlap{\raise.5ex\hbox{$<$}}
                    {\lower.5ex\hbox{$\sim$}}}}

\begin{document}
\preprint{\vbox{\hbox{}\hbox{KEK-TH-927}}}
\draft
\title{An Analysis of $B \to \eta' K$ Decays Using a Global Fit \\
in QCD Factorization}

\author{ Bhaskar Dutta$^{1}$\footnote{duttabh@uregina.ca},
~~ C. S. Kim$^{2}$\footnote{cskim@yonsei.ac.kr},
~~ Sechul Oh$^{3}$\footnote{scoh@post.kek.jp}, ~~ and ~~
Guohuai Zhu$^{3}$\footnote{zhugh@post.kek.jp}  }

\affiliation{
$^1$Department of Physics, University of Regina, SK, S4S 0A2, Canada \\
$^2$Department of Physics, Yonsei University, Seoul
120-479, Korea \\
$^3$Theory Group, KEK, Tsukuba, Ibaraki 305-0801, Japan}

\begin{abstract}
In the framework of QCD factorization, we study $B^{+(0)} \to \eta' K^{+(0)}$
decays.  In order to more reliably determine the phenomenological parameters
$X_H$ and $X_A$ arising from end-point divergences in the hard spectator
scattering and weak annihilation contributions, we use the global analysis
for twelve $B \to PP$ and $VP$ decay modes, such as $B \to \pi \pi$, $\pi K$,
$\rho \pi$, $\rho K$, etc, but excluding the modes whose (dominant) internal
quark-level process is $b \to s \bar s s$.
Based on the global analysis, we critically investigate possible magnitudes
of $X_{H,A}$ and find that both large and small $X_{H,A}$ terms
are allowed by the global fit.  In the case of the large $X_{H,A}$ effects,
the standard model (SM) prediction of the branching ratios (BRs) for
$B^{+(0)} \to \eta' K^{+(0)}$  is
large and well consistent with the experimental results.
In contrast, in the case of the small $X_{H,A}$ effects, the SM prediction
for these BRs is smaller than the experimental data.
Motivated by the recent Belle measurement of $\sin (2\phi_1)$ through
$B^0 \to \phi K_s$, if we take into account possible new physics effects on
the quark-level process $b \to s \bar s s$, we can explicitly show that these
large BRs can be understood even in the small $X_{H,A}$ case.
Specifically, we present two new physics scenarios: R-parity violating SUSY
and R-parity conserving SUSY.
\end{abstract}
\maketitle

\newpage
\section{Introduction}

From $B$ factory experiments such as Belle and BaBar, copious
experimental data on $B$ decays start to provide new bounds on
previously known observables with great precision as well as an
opportunity to see very rare decay modes for the first time. There
exist plenty of experimental data observed for charmless hadronic
decays $B \to PP$ ($P$ denotes a pseudoscalar meson), such as $B
\to \pi \pi$, $\pi K$, etc, and $B \to VP$ ($V$ denotes a vector
meson), such as $B \to \rho \pi$, $\omega \pi$, $\rho K$, etc,
which are well understood within the standard model (SM). However,
among the $B \to PP$ decay modes, the BR of the decay modes
$B^{\pm(0)} \to \eta^{\prime} K^{\pm(0)}$ is found to be still
larger than that expected  within the SM. For last several years
the experimental results of unexpectedly large branching ratios
(BRs) for $B \to \eta^\prime K$ decays have drawn a lot of
theoretical attentions. The observed BRs for $B^{\pm} \to \eta'
K^{\pm}$ in three different experiments are
\cite{Gordon:2002yt,Aubert:2003bq,Cronin-Hennessy:kg}
\begin{eqnarray}
{\mathcal B}( B^{\pm} \to \eta' K^{\pm})
&=& ( 77.9^{+6.2+9.3}_{-5.9-8.7}) \times 10^{-6} ~~[{\rm BELLE}],
\nonumber \\
\mbox{} &=& ( 76.9 \pm 3.5 \pm 4.4 ) \times 10^{-6} ~~[{\rm BABAR}],
\nonumber \\
\mbox{} &=& ( 80^{+10}_{-9} \pm 7 ) \times 10^{-6}  ~~[{\rm CLEO}].
\end{eqnarray}
Many theoretical efforts have been made to explain the large BRs:
for instance, approaches using
the anomalous $g{\mbox -}g{\mbox -}\eta^\prime$ coupling
\cite{Soni,Kagan,Kou1,akoy},
high charm content in $\eta^\prime$ \cite{charm,Ko,ddo},
the spectator hard scattering mechanism \cite{Du,Yang},
the QCD factorization (QCDF) approach \cite{Beneke:2002jn},
the perturbative QCD (PQCD) approach \cite{pQCD} and
approaches to invoke new physics
\cite{Choudhury:1998wc,dko1,dko2,Kundu,Xiao}.

In earlier works on nonleptonic decays of $B$ mesons, the factorization
approximation, based on the color transparency argument, was usually
assumed to estimate the hadronic matrix elements which are inevitably
involved in theoretical calculations of the decay amplitudes for these
processes.  This naive factorization approach ignores the nonfactorizable
contributions from the soft interactions in the initial and final states.
In order to compensate the nonfactorizable contributions, the naive
factorization scheme has been generalized by introducing the effective
number of colors $N_c$ as a phenomenological parameter.
In this generalized factorization, the renormalization scheme and scale
dependence in the hadronic matrix elements has been resolved
\cite{improvedwc}.

Theoretically, the QCDF approach has provided a novel method to study
nonleptonic $B$ decays. In this approach, the naive factorization
contributions become the leading term and as sub-leading contributions,
radiative corrections from hard gluon exchange can be systematically
calculated by using the perturbative QCD method in the heavy quark limit,
where suppressed power corrections of ${\cal O}(\Lambda_{QCD} /m_b)$ are
neglected.
Since the nonfactorizable contributions in the naive factorization, such
as the contributions from hard scattering with the spectator quark in
the $B$ meson and the contributions from weak annihilation, can be
perturbatively computed, the phenomenological parameter $N_c$ used in
the generalized factorization scheme is no longer needed to compensate
the nonfactorizable contributions.

However, in reality the $b$ quark is not very heavy so that the power
corrections in $1/m_b$, particularly the chirally enhanced corrections,
would not be negligible.  The chirally enhanced corrections come from
twist-3 light cone distribution amplitudes (LCDAs), but unfortunately
the QCDF breaks down at twist-3 level because a logarithmic divergence
appears in the hard spectator scattering at the end-point of the
twist-3 LCDAs.  A similar divergence also appears in the weak
annihilation contributions.  It is customary to phenomenologically
treat these two end-point divergences by introducing model-dependent
parameters \cite{Beneke:2001ev}:
$X_H$ for the hard spectator scattering contributions and $X_A$ for
the weak annihilation contributions.
Thus, it would be a less reliable case if these nonpertubative
contributions of $X_H$ and $X_A$ become too large compared with the
leading power radiative corrections.
Since the prediction of the BRs for $B \to PP$ and $B \to VP$ decays
strongly depend on the parameters $X_H$ and $X_A$, it is essential
to reliably estimate the effects of $X_H$ and $X_A$.

In this work we study the decay processes $B^{\pm (0)} \to \eta'
K^{\pm (0)}$ in the QCDF approach.
In order to determine the parameters $X_H$ and $X_A$ more reliably,
we use the global analysis as used in Ref. \cite{Du:2002cf}.
However, our global analysis differs from that used in \cite{Du:2002cf},
in the sense that we exclude the decay modes
whose (dominant) internal quark-level process is $b \to s \bar s s$:
for example, $B \to \phi K$ and $B \to \eta^{(\prime)} M$, where $M$
denotes a light meson, such as $\pi, ~ K, ~ \rho, ~ K^*$.
The reason for excluding such modes is that the recent Belle measurement
of the large negative value of $\sin(2 \phi_1)_{\phi K_s}$ ($\phi_1$
is the angle of the unitarity triangle) through the time dependent decay
process $B^0 \to \phi K_s$ shows a possibility that there may be new physics
effects on the quark-level process $b \to s \bar s s$ \cite{Abe:2003yt}.
Thus, to be conservative, in our global analysis within the SM, all the
decay channels whose (dominant) quark-level process is $b \to s \bar s s$
are excluded so that parameters $X_H$ and $X_A$ can be determined without
new physics prejudice when using the global fit.
For the analysis, we will use twelve $B \to PP$ and $VP$ decay modes,
including $B \to \pi \pi$, $\pi K$, $\rho \pi$, $\rho K$, $\omega \pi$,
$\omega K$.
It turns out that both cases of the large and small $X_{H,A}$ effects
are allowed by the global fit.  We will take into account both
possibilities.  In particular, motivated by the recent Belle result on
$\sin (2 \phi_1)_{\phi K_s}$,
we will seriously examine new physics effects on the large BRs for
$B \to \eta' K$.  As specific examples of new physics models, we will
present both R-parity violating (RPV) supersymmetry (SUSY) and R-parity
conserving (RPC) SUSY scenario.

This work is organized as follows.
In Sec. II, we introduce the framework: the effective Hamiltonian for
nonleptonic charmless $B$ decays and the QCDF approach.
The decay amplitudes for $B \to \eta^{(\prime)} K$ in the QCDF are
presented in Sec. III. In Sec. IV, we discuss the global analysis for
$B \to PP$ and $VP$ decays and calculate the BRs for $B \to \eta' K$
decays as well as $B \to \phi K$ by using the inputs determined from the
global analysis.  We present the results for both cases of the large and
small $X_{H,A}$ effects.  In particular, in the case of the small
$X_{H,A}$ effects, two new physics scenarios (RPV SUSY and RPC SUSY)
are considered.  We conclude the analysis in Sec. V.

\section{Framework}

The effective Hamiltonian for hadronic charmless $B$ decays can be
written as
\begin{eqnarray}
 {\cal H}_{eff}&=&\frac{G_{F}}{\sqrt{2}}
 \Big\{ {\sum\limits_{p=u,c}} V_{pb} V^*_{pq}
     \Big[ C_{1}({\mu}) O^{p}_{1}({\mu})
         + C_{2}({\mu}) O^{p}_{2}({\mu})
         +{\sum\limits_{k=3}^{10}} C_{k}({\mu}) O_{k}({\mu})
     \Big] \nonumber \\
 & & - V_{tb} V^*_{tq} \Big[ C_{7{\gamma}} O_{7{\gamma}}
                + C_{8g} O_{8g} \Big] \Big\} + H.c. , ~~~ (q = d,s)
\end{eqnarray}
where the dimension-6 local operators $O_i$ are given by
\begin{eqnarray}
  & &O^{u}_{1}=({\bar{u}}_{\alpha}b_{\alpha})_{V-A}
               ({\bar{q}}_{\beta} u_{\beta} )_{V-A},
     \ \ \ \ \ \ \ \ \ \ \ \ \ \ \ \ \ \ \
     O^{c}_{1}=({\bar{c}}_{\alpha}b_{\alpha})_{V-A}
               ({\bar{q}}_{\beta} c_{\beta} )_{V-A},
 \nonumber \\
  & &O^{u}_{2}=({\bar{u}}_{\alpha}b_{\beta} )_{V-A}
               ({\bar{q}}_{\beta} u_{\alpha})_{V-A},
     \ \ \ \ \ \ \ \ \ \ \ \ \ \ \ \ \ \ \
     O^{c}_{2}=({\bar{c}}_{\alpha}b_{\beta} )_{V-A}
            ({\bar{q}}_{\beta} c_{\alpha})_{V-A},
 \nonumber \\
  & &O_{3}=({\bar{q}}_{\alpha}b_{\alpha})_{V-A}\sum\limits_{q^{\prime}}
           ({\bar{q}}^{\prime}_{\beta} q^{\prime}_{\beta} )_{V-A},
     \ \ \ \ \ \ \ \ \ \ \ \ \ \ \ \
     O_{4}=({\bar{q}}_{\beta} b_{\alpha})_{V-A}\sum\limits_{q^{\prime}}
           ({\bar{q}}^{\prime}_{\alpha}q^{\prime}_{\beta} )_{V-A},
 \nonumber \\
  & &O_{5}=({\bar{q}}_{\alpha}b_{\alpha})_{V-A}\sum\limits_{q^{\prime}}
           ({\bar{q}}^{\prime}_{\beta} q^{\prime}_{\beta} )_{V+A},
     \ \ \ \ \ \ \ \ \ \ \ \ \ \ \ \
     O_{6}=({\bar{q}}_{\beta} b_{\alpha})_{V-A}\sum\limits_{q^{\prime}}
           ({\bar{q}}^{\prime}_{\alpha}q^{\prime}_{\beta} )_{V+A},
 \nonumber \\
  & &O_{7}=\frac{3}{2}({\bar{q}}_{\alpha}b_{\alpha})_{V-A}
           \sum\limits_{q^{\prime}}e_{q^{\prime}}
           ({\bar{q}}^{\prime}_{\beta} q^{\prime}_{\beta} )_{V+A},
     \ \ \ \ \ \ \ \ \ \
     O_{8}=\frac{3}{2}({\bar{q}}_{\beta} b_{\alpha})_{V-A}
           \sum\limits_{q^{\prime}}e_{q^{\prime}}
           ({\bar{q}}^{\prime}_{\alpha}q^{\prime}_{\beta} )_{V+A},
 \nonumber \\
  & &O_{9}=\frac{3}{2}({\bar{q}}_{\alpha}b_{\alpha})_{V-A}
           \sum\limits_{q^{\prime}}e_{q^{\prime}}
           ({\bar{q}}^{\prime}_{\beta} q^{\prime}_{\beta} )_{V-A},
     \ \ \ \ \ \ \ \ \ \
    O_{10}=\frac{3}{2}({\bar{q}}_{\beta} b_{\alpha})_{V-A}
           \sum\limits_{q^{\prime}}e_{q^{\prime}}
           ({\bar{q}}^{\prime}_{\alpha}q^{\prime}_{\beta} )_{V-A},
 \nonumber \\
  & &O_{7{\gamma}}=\frac{e}{8{\pi}^{2}}m_{b}{\bar{q}}_{\alpha}
           {\sigma}^{{\mu}{\nu}}(1+{\gamma}_{5})
            b_{\alpha}F_{{\mu}{\nu}},
     \ \ \ \ \ \ \ \ \ \ \
     O_{8g}=\frac{g}{8{\pi}^{2}}m_{b}{\bar{q}}_{\alpha}
           {\sigma}^{{\mu}{\nu}}(1+{\gamma}_{5})
            t^{a}_{{\alpha}{\beta}}b_{\beta}G^{a}_{{\mu}{\nu}},
\end{eqnarray}
where $q^{\prime}$ denotes all the active quarks at the scale
${\mu}={\cal O}(m_{b})$, i.e., $q^{\prime}=u,d,s,c,b$.
The operators $O^p_1$, $O^p_2$ are the tree operators, $O_{3-6}$ are the
strong penguin operators, $O_{7-10}$ are the electroweak penguin operators,
and $O_{7 \gamma}$, $O_{8 g}$ are the magnetic penguin operators.
The Wilson coefficients (WCs) $C_{i}({\mu})$ are obtained by running the
renormalization group equations from the weak scale down to scale $\mu$.
We will use the WCs evaluated to the next-to-leading logarithmic order in
the NDR scheme, as given in Ref. \cite{Du:2001hr}.

In the QCDF approach, in the heavy quark limit $m_b >> \Lambda_{\rm QCD}$,
the hadronic matrix element for $B \to M_1 M_2$ due to a particular
operator $O_i$ can be written in the form
\begin{eqnarray}
\langle M_1 M_2 | O_i | B \rangle
 = \langle M_1 M_2 | O_i | B \rangle_{\rm NF}
 \cdot \left[ 1 + \sum_n r_n (\alpha_s)^n
 + {\cal O} \left( {\Lambda_{\rm QCD} \over m_b} \right) \right],
\end{eqnarray}
where $\langle M_1 M_2 | O_{i} | B \rangle_{\rm NF}$ denotes the naive
factorization result.  The second and third term in the square bracket
represent the radiative corrections in $\alpha_s$ and the power
corrections in $\Lambda_{\rm QCD} / m_b$.
The decay amplitudes for $B \to M_1 M_2$ can be expressed as
\begin{eqnarray}
 {\cal A}(B \to M_1 M_2) = {\cal A}^f(B \to M_1 M_2)
 + {\cal A}^a(B \to M_1 M_2)~,
\end{eqnarray}
where
\begin{eqnarray}
 {\cal A}^f (B \to M_1 M_2) &=& \frac{G_{F}}{\sqrt{2}}
    \sum\limits_{p=u,c} \sum\limits_{i=1}^{10} V_{pb} V^*_{pq}
    ~ a_{i}^{p} \langle M_1 M_2 {\vert} O_{i} {\vert} B
    \rangle_{\rm NF}~, \nonumber \\
 {\cal A}^a (B \to M_1 M_2) &=& \frac{G_{F}}{\sqrt{2}}~
    f_{B} f_{M_1} f_{M_2} \sum\limits_{p=u,c} \sum\limits_{i=1}^{10}
    V_{pb} V^*_{pq}~ b_{i} ~.
\end{eqnarray}
Here ${\cal A}^f (B \to M_1 M_2)$ includes vertex corrections, penguin
corrections, and hard spectator scattering contributions which are
absorbed into the QCD coefficients $a_i$, and ${\cal A}^a (B \to M_1 M_2)$
includes weak annihilation contributions which are absorbed into the
parameter $b_i$.  For the explicit expressions of $a_i$ and $b_i$, we
refer to Refs. \cite{Beneke:2001ev,Du:2001hr}.

It is well known \cite{Beneke:2001ev} that both in the hard spectator
scattering and in the annihilation contributions there appears logarithmic
divergence in the end-point region.  In Ref. \cite{Beneke:2001ev},
Beneke {\it et al.} introduced phenomenological parameters for the
end-point divergent integrals:
\begin{eqnarray}
X_{H,A} \equiv \int^1_0 {dx \over x}
 \equiv \left( 1 + \rho_{H,A} e^{i \phi_{H,A}} \right)
 {\rm ln} {m_B \over \Lambda_h} ~,
\label{XAXH}
\end{eqnarray}
where $X_H$ and $X_A$ denote the hard spectator scattering contribution and
the annihilation contribution, respectively.
Here the phases $\phi_{H,A}$ are arbitrary, $0^0 \leq \phi_{H,A} \leq 360^0$,
and the parameter $\rho_{H,A} \leq 1$ and the scale
$\Lambda_h = 0.5$ GeV assumed phenomenologically \cite{Beneke:2001ev}.
In principle, the parameters $\rho_{H,A}$ and $\phi_{H,A}$ for $B \to PP$
decays can be different from those for $B \to VP$ decays.
Thus, for $B \to PP$ and $VP$ decays,  from the end-point divergent
integrals, eight new parameters are introduced: $\rho^{PP}_{H,A}, ~
\phi^{PP}_{H,A}$ for $B \to PP$, and $\rho^{VP}_{H,A},~ \phi^{VP}_{H,A}$
for $B \to VP$.

\section{Decay processes $B^{\pm (0)} \to \eta^{(\prime)} K^{\pm (0)}$
in the QCDF approach}

The decay amplitudes for $B^- \to \eta^{(\prime)} K^-$ and
$\bar{B^0} \to \eta^{(\prime)} \bar{K^0}$ in the QCDF are given by
\begin{eqnarray}
{\cal A}(B^- \to \eta^{(\prime)} K^-) =
&-& i {G_F \over \sqrt{2}}
 f_K F_0^{B \to \eta^{(\prime)}}(m_K^2)(m_B^2 -m_{\eta^{(\prime)}}^2)
\nonumber \\
&\mbox{}& \cdot
 [ V_{ub} V^*_{us} ( a_1^{\prime} +a_4^{u ~ \prime}
  +a_{10}^{u ~ \prime} +(a_6^{u ~ \prime} +a_8^{u ~ \prime}) R_1)
\nonumber \\
&\mbox{}&
  + V_{cb} V^*_{cs} ( a_4^{c ~ \prime} +a_{10}^{c ~ \prime}
  +(a_6^{c ~ \prime} +a_8^{c ~ \prime}) R_1) ]
\nonumber \\
&-& i {G_F \over \sqrt{2}} F_0^{B \to K}(m_{\eta^{(\prime)}}^2)(m_B^2 -m_K^2)
\nonumber \\
&\mbox{}& \cdot \left\{ f_{\eta^{(\prime)}}^u \left[ V_{ub} V^*_{us}
  \left( a_2 +2 a_3 -2 a_5 -{1 \over 2} (a_7 -a_9 )
   - \left( a_6^u -{1 \over 2} a_8^u \right) R_3 \right) \right. \right.
\nonumber \\
&\mbox{}& \left. \left. + V_{cb} V^*_{cs}
  \left(2 a_3 -2 a_5 -{1 \over 2}(a_7 -a_9)
   - \left( a_6^c -{1 \over 2} a_8^c \right) R_3 \right) \right] \right.
\nonumber \\
&\mbox{}& + f_{\eta^{(\prime)}}^s \left[ V_{ub} V^*_{us}
  \left(a_3 +a_4^u -a_5 +{1 \over 2}(a_7 -a_9 -a_{10})
   + \left( a_6^u -{1 \over 2} a_8^u \right) R_3 \right) \right.
\nonumber \\
&\mbox{}& + \left. \left. V_{cb} V^*_{cs}
  \left(a_3 +a_4^c -a_5 +{1 \over 2}(a_7 -a_9 -a_{10})
   + \left( a_6^c -{1 \over 2} a_8^c \right) R_3 \right) \right] \right\}
\nonumber \\
&-& i {G_F \over \sqrt{2}} f_B f_K \left( f^u_{\eta^{(\prime)}}
 +f^s_{\eta'} \right)
 ~ \left[ V_{ub} V^*_{us} b_2
  + \left( V_{ub} V^*_{us} +V_{cb} V^*_{cs} \right) (b_3 +b_3^{ew}) \right]~,
\label{charged}  \\
{\cal A}(\bar{B^0} \to \eta^{(\prime)} \bar{K^0}) =
&-& i {G_F \over \sqrt{2}}
 f_K F_0^{B \to \eta^{(\prime)}}(m_K^2)(m_B^2 -m_{\eta^{(\prime)}}^2)
\nonumber \\
&\mbox{}& \cdot
 [ V_{ub} V^*_{us} (a_4^{u ~ \prime} -{1 \over 2} a_{10}^{u ~ \prime}
  +(a_6^{u ~ \prime} -{1 \over 2} a_8^{u ~ \prime}) R_2)
\nonumber \\
&\mbox{}&
  + [ V_{cb} V^*_{cs} (a_4^{c ~ \prime} -{1 \over 2} a_{10}^{c ~ \prime}
  +(a_6^{c ~ \prime} -{1 \over 2} a_8^{c ~ \prime}) R_2)
\nonumber \\
&-& i {G_F \over \sqrt{2}} F_0^{B \to K}(m_{\eta'}^2)(m_B^2 -m_K^2)
\nonumber \\
&\mbox{}& \cdot \left\{ f_{\eta^{(\prime)}}^u \left[ V_{ub} V^*_{us}
  \left( a_2 +2 a_3 -2 a_5 -{1 \over 2}(a_7 -a_9)
   - \left( a_6^u -{1 \over 2} a_8^u \right) R_3 \right) \right. \right.
\nonumber \\
&\mbox{}& \left. \left. + V_{cb} V^*_{cs}
  \left(2 a_3 -2 a_5 -{1 \over 2}(a_7 -a_9)
   - \left( a_6^c -{1 \over 2} a_8^c \right) R_3 \right) \right] \right.
\nonumber \\
&\mbox{}& + f_{\eta^{(\prime)}}^s \left[ V_{ub} V^*_{us}
  \left(a_3 +a_4^u -a_5 +{1 \over 2}(a_7 -a_9 -a_{10})
   + \left( a_6^u -{1 \over 2} a_8^u \right) R_3 \right) \right.
\nonumber \\
&\mbox{}& + \left. \left. V_{cb} V^*_{cs}
  \left(a_3 +a_4^c -a_5 +{1 \over 2}(a_7 -a_9 -a_{10})
   + \left( a_6^c -{1 \over 2} a_8^c \right) R_3 \right) \right] \right\}
\nonumber \\
&-& i {G_F \over \sqrt{2}} f_B f_K \left( f^u_{\eta^{(\prime)}}
 +f^s_{\eta^{(\prime)}} \right)
 ~\left( V_{ub} V^*_{us} +V_{cb} V^*_{cs} \right)
  \left( b_3 -{1 \over 2} b_3^{ew} \right) ~,
\label{neutral}
\end{eqnarray}
where
\begin{eqnarray}
R_{1(2)} = {2 m_{K^{(0)}}^2 \over (m_b -m_{u(d)}) (m_{u(d)} +m_s)} ~, ~~~
R_3 = {2 m_{\eta'}^2 \over 2 m_s (m_b -m_s)}~.
\end{eqnarray}
The coefficients $a^{(\prime)}_i$ and $b_i$ are expressed as
\begin{eqnarray}
a^{(\prime)}_1 &=& C_1 +{C_2 \over N_c} \left[ 1 + {C_F \alpha_s \over 4 \pi}
 \left( V_M +{4 \pi^2 \over N_c} H(B M_1, M_2) \right) \right], \nonumber \\
a_2 &=& C_2 +{C_1 \over N_c} \left[ 1 + {C_F \alpha_s \over 4 \pi}
 \left( V_M +{4 \pi^2 \over N_c} H(B M_1, M_2) \right) \right], \nonumber \\
a_3 &=& C_3 +{C_4 \over N_c} \left[ 1 + {C_F \alpha_s \over 4 \pi}
 \left( V_M +{4 \pi^2 \over N_c} H(B M_1, M_2) \right) \right], \nonumber \\
a^{p~(\prime)}_4 &=& C_4 +{C_3 \over N_c} \left[ 1 + {C_F \alpha_s \over 4 \pi}
 \left( V_M +{4 \pi^2 \over N_c} H(B M_1, M_2) \right) \right]
 + {C_F \alpha_s \over 4 \pi N_c} P_{M,2}^p, \nonumber \\
a_5 &=& C_5 +{C_6 \over N_c} \left[ 1 - {C_F \alpha_s \over 4 \pi}
 \left( V_M +12 +{4 \pi^2 \over N_c} H(B M_1, M_2) \right) \right],
 \nonumber \\
a^{p~(\prime)}_6 &=& C_6 +{C_5 \over N_c} \left[ 1 - 6{C_F \alpha_s \over 4 \pi}
 \right] + {C_F \alpha_s \over 4 \pi N_c} P_{M,3}^p, \nonumber \\
a_7 &=& C_7 +{C_8 \over N_c} \left[ 1 - {C_F \alpha_s \over 4 \pi}
 \left( V_M +12 +{4 \pi^2 \over N_c} H(B M_1, M_2) \right) \right],
 \nonumber \\
a^{p~(\prime)}_8 &=& C_8 +{C_7 \over N_c} \left[ 1 - 6{C_F \alpha_s \over 4 \pi}
 \right] + {\alpha \over 9 \pi N_c} P_{M,3}^{p,ew}, \nonumber \\
a_9 &=& C_9 +{C_{10} \over N_c} \left[ 1 + {C_F \alpha_s \over 4 \pi}
 \left( V_M +{4 \pi^2 \over N_c} H(B M_1, M_2) \right) \right], \nonumber \\
a^{p~(\prime)}_{10} &=& C_{10} +{C_9 \over N_c} \left[ 1
 + {C_F \alpha_s \over 4 \pi}
 \left( V_M +{4 \pi^2 \over N_c} H(B M_1, M_2) \right) \right]
 + {\alpha \over 9 \pi N_c} P_{M,2}^{p,ew}, \nonumber \\
b_2 &=& {C_F \over N_c^2} C_2 A^i, \nonumber \\
b_3 &=& {C_F \over N_c^2} [ C_3 A^i + A^f (C_5 +N_c C_6)],  \nonumber \\
b_3^{ew} &=& {C_F \over N_c^2} [ C_9 A^i + A^f (C_7 +N_c C_8)],
\end{eqnarray}
where the superscript $p$ is $u$ or $c$, and the color factor
$C_F = (N_c^2 -1)/ (2 N_c)$ with $N_c =3$.
The vertex parameter $V_M$ and the hard spectator scattering parameter
$H(B M_1, M_2)$, and the weak annihilation parameters $A^i, ~ A^f$  are given
by \cite{Beneke:2001ev,Du:2001hr}
\begin{eqnarray}
V_M &=& 12 \ln{m_b \over \mu} -18 + \int_0^1 dx ~ g(x) \Phi_M (x), \nonumber \\
H(B M_1, M_2) &=& {f_B f_{M_1} \over m_B^2 F_0^{B \to M_1}}
 \int_0^1 d\xi {\Phi_B (\xi) \over \xi} \int_0^1 dx {\Phi_{M_2} (x) \over (1-x)}
\nonumber \\
&\mbox{}& \times
 \int_0^1 dy \left[ {\Phi_{M_1} (y) \over (1-y)}
 +{2 \mu_{M_1} \over m_b} {(1-x) \over x} {\Phi^p_{M_1} (y) \over (1-y)}
 \right],  \nonumber \\
A^i &\approx& \pi \alpha_s \left[ 18 \left( X_A -4 +{\pi^2 \over 3} \right)
 +2 r_{\chi}^2 X_A^2 \right],  \nonumber \\
A^f &\approx& 12 \pi \alpha_s r_{\chi} X_A (2 X_A -1),
\label{VHAs}
\end{eqnarray}
where $g(x) =3 \left( {1 -2x \over 1 -x} \right) \ln{x} -3i \pi$ and
$\mu_P = {m_P^2 \over m_1 + m_2}$ ($m_1$ and $m_2$ are current quark masses
of the valence quarks of the meson $P$) and the chirally enhanced factor
$r_{\chi}= {2 \mu_P \over m_b}$.
For the chirally enhanced parameter $r_{\chi}$, we will take
$r^{\eta'}_{\chi} \left( 1 -{f^u_{\eta'} \over f^s_{\eta'}} \right)
= r^{\pi}_{\chi} = r^K_{\chi} \equiv r_{\chi}$ as in Ref. \cite{Du:2001hr}.
$X_A \equiv \int_0^1 {dx \over x}$ is a logarithmically divergent integral.
For the wave function $\Phi_B (\xi)$ of the $B$ meson, we take the following
parametrization:
\begin{eqnarray}
\int_0^1 d\xi ~ {\Phi_B (\xi) \over \xi} \equiv {m_B \over \lambda_B},
\end{eqnarray}
where the parameter $\lambda_B$ is estimated as $\lambda_B = (350 \pm 150)
~{\rm MeV}$ \cite{Beneke:2001ev}.
For the $K$ and $\eta'$ meson, we use the asymptotic forms of the LCDAs
\cite{Beneke:2001ev}:
\begin{eqnarray}
\Phi_K (x) &=& \Phi_{\eta'} = 6 x (1-x),  \nonumber \\
\Phi_K^p (x) &=& \Phi_{\eta'}^p (x) = 1,
\end{eqnarray}
where $\Phi_M (x)$ and $\Phi_M^p (x)$ are the leading twist LCDAs and
twist-3 LCDAs of the meson $M = K, ~ \eta'$, respectively.
The explicit expressions of the QCD penguin parameters $P^p_{M,i}$ and
the electroweak penguin parameters $P^{p,ew}_{M,i}$ can be found in
Refs. \cite{Beneke:2001ev,Du:2001hr}.
The coefficients $a_i$ and $a_i^{\prime}$ in Eqs. (\ref{charged})
and (\ref{neutral}) include the different vertex and hard spectator
scattering contributions:
for $a_i$, $V_M = V_{\eta'}$ and $H(B M_1, M_2) = H(B K, \eta')$,
while for $a_i^{\prime}$, $V_M = V_K$ and $H(B M_1, M_2) = H(B \eta', K)$.

Note that in Eq. (\ref{VHAs}) the hard spectator scattering parameter
$H(B M_1, M_2)$ includes a logarithmically divergent integral
$\int_0^1 dy / (1-y)$ which arises from the twist-3 contribution, and the
weak annihilation parameters $A^i$ and $A^f$ include another logarithmically
divergent integral $X_A$.

For the $\eta - \eta'$ mixing, we use the following relation:
\begin{eqnarray}
| \eta \rangle &=& \cos \theta_8 | \eta_8 \rangle -\sin \theta_0
| \eta_0 \rangle , \nonumber \\
| \eta' \rangle &=& \sin \theta_8 | \eta_8 \rangle +\cos \theta_0
| \eta_0 \rangle ,
\end{eqnarray}
where $\eta_8$ and $\eta_0$ are the flavor SU(3) octet and single,
respectively. The mixing angles are $\theta_8 \approx -22.2^0$ and
$\theta_0 \approx -9.1^0$ \cite{Feldmann:1998vh}.  The decay constants
and form factors relevant for the $B \to \eta^{(\prime)}$ transitions
are given by
\begin{eqnarray}
f^u_{\eta} &=& {f_8 \over \sqrt{6}} \cos\theta_8
 -{f_0 \over \sqrt{3}} \sin\theta_0 , ~~
f^s_{\eta} = -2 {f_8 \over \sqrt{6}} \cos\theta_8
 -{f_0 \over \sqrt{3}} \sin\theta_0 ,  \nonumber \\
f^u_{\eta'} &=& {f_8 \over \sqrt{6}} \sin\theta_8
 +{f_0 \over \sqrt{3}} \cos\theta_0 , ~~
f^s_{\eta'} = -2 {f_8 \over \sqrt{6}} \sin\theta_8
 +{f_0 \over \sqrt{3}} \cos\theta_0 ,  \nonumber \\
F^{B \eta}_{0,1} &=& F^{B \pi}_{0,1} \left( {\cos\theta_8 \over \sqrt{6}}
 -{\sin\theta_8 \over \sqrt{3}} \right),  ~~
F^{B \eta'}_{0,1} = F^{B \pi}_{0,1} \left( {\sin\theta_8 \over \sqrt{6}}
 +{\cos\theta_8 \over \sqrt{3}} \right).
\end{eqnarray}

\section{Global Analysis and Numerical Result}

In order to calculate the BRs for $B$ decays in the QCDF approach, various
input parameters are needed, such as the CKM matrix elements, decay constants,
transition form factors, LCDAs, and so on.  Among those input parameters,
it is urgently essential to reliably estimate the annihilation parameter $X_A$
and the hard spectator scattering parameter $X_H$: more specifically, $\rho_A$,
$\phi_A$, $\rho_H$, and $\phi_H$, because the predicted BRs strongly depend
on the parameters $X_A$ and $X_H$ (see Figs. 1 and 2).
Unfortunately, within the QCDF scheme, $X_A$ and $X_H$ are purely
phenomenological parameters, so there is no definite way to determine
them.  Therefore, in order to determine the values of $\rho_A$, $\phi_A$,
$\rho_H$, and $\phi_H$ more reliably, in this work we follow the global analysis,
used in Ref. \cite{Du:2002cf}. For the detailed discussion on the method of
the global fit, we refer to Ref. \cite{Du:2002cf}.
As explained in Sec. I, different from the global analysis used in
\cite{Du:2002cf}, we do not include the decay modes, such as $B \to \phi K$
and $B \to \eta^{(\prime)} M$ ($M$ denotes a light meson: e.g., $\pi$, $K$,
$\rho$, $K^*$), whose (dominant) internal quark-level process is
$b \to s \bar s s$.
In this way, within the SM, the parameters $\rho_A$, $\phi_A$, $\rho_H$,
$\phi_H$ can be determined without new physics prejudice
when using the global fit.
Specifically, we use twelve decay modes, such as $B \to \pi \pi$, $\pi K$,
$\rho \pi$, $\rho K$, $\omega \pi$, and $\omega K$, as listed in Table I
\cite{Gordon:2002yt,Aubert:2003bq,Cronin-Hennessy:kg}.

First, we examine the dependence of the BR for $B^+ \to \eta' K^+$ on the
effects of $X_A$ and $X_H$.  Figure 1 shows ${\cal B}(B^+ \to \eta' K^+)$
versus $\phi^{PP}_A$ (solid line) or $\phi^{PP}_H$ (dotted line).
In each case, $\phi^{PP}_A$ or $\phi^{PP}_H$ varies from 0 to $2\pi$.
For the solid line, other inputs are set as $\rho^{PP}_A =\rho^{PP}_H =1$,
$\phi^{PP}_H =-23^0$.  For the dotted line, $\rho^{PP}_A =\rho^{PP}_H =1$,
$\phi^{PP}_A =57^0$.  We see that the predicted BR for $B^+ \to \eta' K^+$
strongly depends on $\phi^{PP}_A$ and $\phi^{PP}_H$.  In particular, as
the value of $\phi^{PP}_A$ varies, the predicted BR can change by a factor
of about 2.5 (e.g., from $45 \times 10^{-6}$ to $116 \times 10^{-6}$).
The allowed values of $\phi^{PP}_A$ are in certain narrow regions which can
be practically found by the global analysis.
Similarly, Figure 2 shows ${\cal B}(B^+ \to \eta' K^+)$ versus $\rho^{PP}_A$
(solid line) or $\rho^{PP}_H$ (dotted line). In each case, $\rho^{PP}_A$
or $\rho^{PP}_H$ varies from 0 to 1.
The other inputs are put as $\phi^{PP}_A = -23^0$ and $\phi^{PP}_H =57^0$
for both lines, $\rho^{PP}_H =1$ for the solid line, and $\rho^{PP}_A =1$
for the dotted line.
The predicted BR for $B^+ \to \eta' K^+$ is also dependent on $\rho^{PP}_A$
and $\rho^{PP}_H$, but its dependence on $\rho^{PP}_{A,H}$ is weaker
than that on $\phi^{PP}_A$.
We notice that the prediction of ${\cal B}(B^+ \to \eta' K^+)$ is very
sensitive to the effect of $X_A$ through $\phi^{PP}_A$.  This feature
also holds for the neutral mode $B^0 \to \eta' K^0$.

\begin{figure}
    \centerline{ \DESepsf(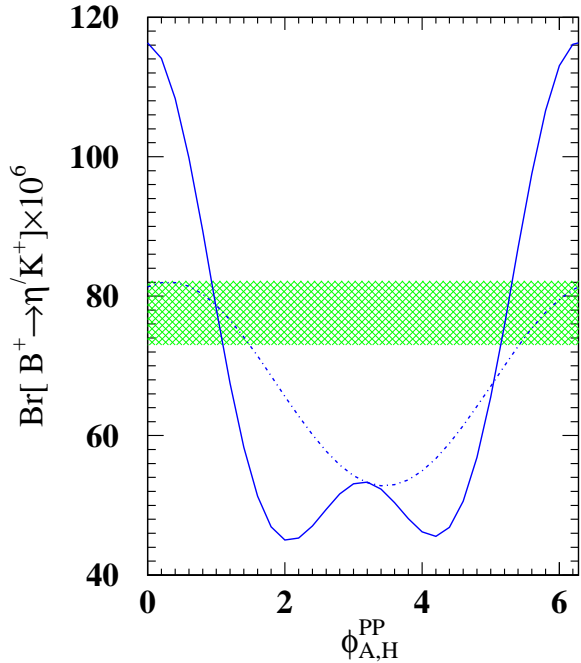 width 7cm)}
    \vspace{0.1cm}
    \caption{\label{fig:fig1}  Dependence of the BR for $B^+ \to \eta' K^+$
    on $\phi^{PP}_A$ (solid line) or $\phi^{PP}_H$ (dotted line). Here the
    following values of the other parameters are used:
    $\rho^{PP}_A =\rho^{PP}_H =1$ (for both lines), $\phi^{PP}_H =-23^0$
    (for the solid line), $\phi^{PP}_A =57^0$ (for the dotted line).
    The shaded region is allowed by the experimental data. }
\end{figure}

We find that the best fit [and also the ``good'' fit (see the
discussions below {\bf [Case 1]})] of the global analysis favors
large effects of the parameters $X_A$ and $X_H$. This tendency is
consistent with the results of other previous works done in the
QCDF scheme \cite{Du:2002cf,Beneke:2003zv}. But, as mentioned in
Sec. I, if the nonperturbative effects of $X_A$ and $X_H$ are too
large or dominant compared with the leading power radiative
corrections, the theoretical predictions based on these effects
would be less reliable and become questionable. Therefore, one can
seriously ask the following question: Is it possible to find a
global fit where the effects of $X_A$ and $X_H$ are rather small
(so the theoretical predictions based on these effects would be
more reliable), but its $\chi^2$ value is still acceptably small?
In fact, it turns out that such an acceptable fit with the small
effects of $X_A$ and $X_H$ can be found.

\begin{figure}
    \centerline{ \DESepsf(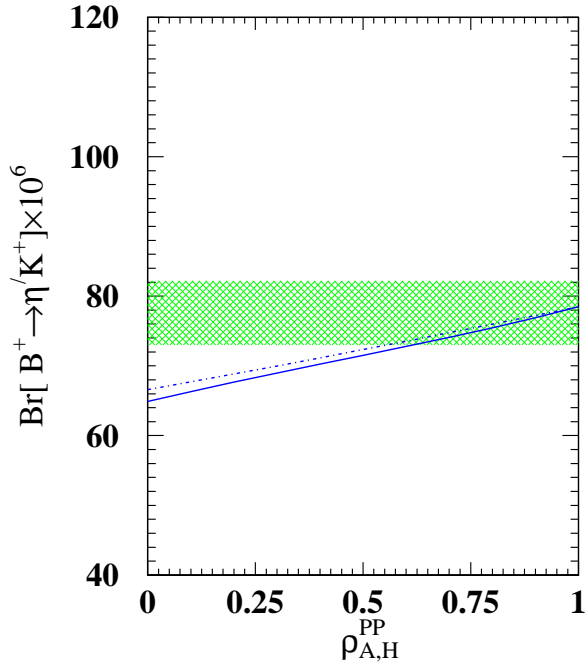 width 7cm)}
    \vspace{0.1cm}
    \caption{\label{fig:fig2}  Dependence of the BR for $B^+ \to \eta' K^+$
    on $\rho^{PP}_A$ (solid line) or $\rho^{PP}_H$ (dotted line). Here the
    following values of the other parameters are used:
    $\phi^{PP}_A = -23^0$ and $\phi^{PP}_H =57^0$ (for both lines),
    $\rho^{PP}_H =1$ (for the solid line), $\rho^{PP}_A =1$
    (for the dotted line).  The shaded region is allowed by the experimental
    data.}
\end{figure}

\begin{table}
\caption{Experimental data and the ``good'' fit values of CP-averaged
branching ratios (in unit of $10^{-6}$) for $B \to PP$ and $VP$ decays
used in our global analysis. }
\smallskip
\begin{tabular}{c|c|c||c|c|c}
\hline \hline
Decay mode & Weighted average(Exp.) & Fit  & Decay mode
& Weighted average(Exp.) & Fit
\\ \hline
   $B^+ \to \pi^+ \pi^0 $ & $5.42 \pm 0.83$ & 4.85
 & $B^0 \to \pi^+ \pi^- $ & $4.55 \pm 0.44$ & 4.75
\\ $B^+ \to \pi^+ K^0$    & $20.8 \pm 1.4$  & 20.5
 & $B^0 \to \pi^+ K^-$    & $18.1 \pm 0.8$  & 18.9
\\ $B^+ \to \pi^0 K^+$    & $12.7 \pm 1.1$  & 11.4
 & $B^0 \to \pi^0 K^0$    & $11.2 \pm 1.4$  & 8.5
\\ $B^+ \to \pi^+ \rho^0$ & $8.6 \pm 2.0$   & 7.6
 & $B^0 \to \pi^{\pm} \rho^{\mp}$  & $22.7 \pm 2.5$ & 23.5
\\ $B^+ \to \omega \pi^+$ & $6.0 \pm 0.9$   & 6.45
 & $B^0 \to K^+ \rho^-$ & $8.0  \pm 1.7$    & 9.5
\\ $B^+ \to \omega K^+$   & $5.6 \pm 0.9$   & 5.25
 & $B^0 \to \omega K^0$   & $5.3  \pm 1.5$  & 4.45
\\ \hline \hline
\end{tabular}
\end{table}

For calculation of the BRs for $B^{+(0)} \to \eta' K^{+(0)}$, we take
into account two different possibilities as discussed above:
[Case 1] with the large $X_A$ and $X_H$ effects (favored by the best
and ``good'' fit, but less reliable), [Case 2] with the small $X_A$
and $X_H$ effects (more reliable).

{\bf [Case 1] with the large $X_A$ and $X_H$ effects}

We first try to find the best fit of the global analysis for the twelve
$B \to PP$ and $VP$ decay channels shown in Table I.  Our result shows
that based on the theoretical inputs for the best fit
(with $\chi^2_{min}=7.5$), the predicted BR for $B^+ \to \eta' K^+$ is
consistent with the experimental data, but the prediction of
${\cal B}(B^+ \to \phi K^+)$ is too small compared with the data
(for the data, see Table II), as the best fit predicted BRs are
\begin{eqnarray}
{\cal B}(B^+ \to \eta' K^+) &=& 74.7 \times 10^{-6},  \nonumber \\
{\cal B}(B^+ \to \phi K^+) &=& 4.02 \times 10^{-6}.
\end{eqnarray}
It happens because the internal interference between different
contributions ($e.g.$, contributions from the hard spectator
scattering and weak annihilation) to the decay amplitude  for $B^+
\to \eta' K^+$ is quite different from that for $B^+ \to \phi K^+$
(for example, see Table II).
It turns out that it is possible to obtain successful fits to all
$B \to \eta' K$ and $B \to \phi K$ data, if one assumes that there
are new physics effects on the quark-level process $b \to s \bar s s$.
We will discuss this possibility later in SUSY scenarios.

Since the input parameter values for the best fit are not consistent with
the experimental result such as the BR for $B \to \phi K$, we investigate
another possibility that there may exist a ``good'' fit for which the
predictions based on the inputs are consistent with the experimental
measurements including $B \to \eta' K$ and $B \to \phi K$, and whose
$\chi^2_{min}$ value is still quite small. In fact, we find such a
``good'' fit with $\chi^2_{min} =8.6$ for the twelve decay modes.
Notice that this $\chi^2_{min}$ value is not much different from that of
the best fit.  In Table I, we list the ``good'' fit values of the BRs for
the relevant $B \to PP$ and $VP$ decay modes.
The corresponding theoretical inputs are given by
\begin{eqnarray}
&\mbox{}& \lambda =0.2205, ~~  A=0.814, ~~ \phi_3 =72^0, ~~
  |V_{ub}|=3.49 \times 10^{-3},
\nonumber \\
&\mbox{}& \mu =2.1 ~{\rm GeV}, ~~ m_s  (2 ~{\rm GeV}) =85 ~{\rm MeV},
 ~~ f_B =220 ~{\rm MeV}, ~~ \lambda_B = 200 ~{\rm MeV},
\nonumber \\
&\mbox{}& F^{B \pi} =0.23, ~~  R_{\pi K} =1, ~~ A^{B \rho} =0.31,
\nonumber \\
&\mbox{}& \rho^{PP}_A = \rho^{VP}_A = \rho^{PP}_H = \rho^{VP}_H =1,
\nonumber \\
&\mbox{}& \phi^{PP}_A =57^0, ~~ \phi^{VP}_A =52^0, ~~
  \phi^{PP}_H =-23^0, ~~ \phi^{VP}_H =180^0,
\label{goodfitinputs}
\end{eqnarray}
where $\lambda \equiv |V_{us}|$ and $R_{\pi K} \equiv
(f_{\pi} F^{BK}) /(f_K F^{B \pi})$.
The parameter $A$ is defined by $A \lambda^2 = |V_{cb}|$ and
$\phi_3$ is the angle of the unitarity triangle.
$f_B$ is the $B$ meson decay constant, and $F^{B \pi}$ and
$A^{B \rho}$ are the form factors for the transition $B \to \pi$ and
$B \to \rho$, respectively.

\begin{table}
\caption{Experimental data and the prediction of the
branching ratios (in unit of $10^{-6}$) for $B \to \eta' K$ and
$B \to \phi K$ decays.  Here the inputs for the ``good'' fit are
used.  For comparison, the predicted BRs for three cases are also
listed: (i) for $X_A =X_H =0$, (ii) for only $X_A =0$, (iii) for only $X_H =0$. }
\smallskip
\begin{tabular}{c||c|c|c|c|c}
\hline \hline
Decay mode & Exp. data  & Prediction  & $X_A =X_H =0$
& $X_A =0$ only & $X_H =0$ only
\\ \hline
   $B^+ \to \eta' K^+ $ & $77.6 \pm 4.6$ & 78.5 & 52.3 & 64.9 & 66.6
\\ $B^0 \to \eta' K^0 $ & $65.0 \pm 6.0$ & 71.6 & 47.8 & 59.5 & 60.7
\\ $B^+ \to \phi K^+$   & $9.3 \pm 0.7$  & 8.85 & 2.27 & 1.51 & 6.49
\\ $B^0 \to \phi K^0$   & $8.2 \pm 1.1$  & 8.01 & 2.06 & 1.37 & 5.85
\\ \hline \hline
\end{tabular}
\end{table}

Indeed the BRs for $B^{+(0)} \to \eta' K^{+(0)}$ and
$B^{+(0)} \to \phi K^{+(0)}$ calculated by using the above inputs
are in good agreement with the experimental measurements as shown
in Table II.  Therefore, our result shows that the large BRs for
the processes $B^{+(0)} \to \eta' K^{+(0)}$ as well as the BRs for
$B^{+(0)} \to \phi K^{+(0)}$ can be consistently understood,
based on the global analysis for $B \to PP$ and $VP$ decays,
where the values of the pure phenomenological parameters
$\rho^{PP}_{A,H}$, $\rho^{VP}_{A,H}$, $\phi^{PP}_{A,H}$ and
$\phi^{VP}_{A,H}$ are reasonably determined.
The BRs for $B^{+(0)} \to \eta K^{+(0)}$ are estimated as
$(1 \sim 2) \times 10^{-6}$ which are also consistent with the data
\cite{Gordon:2002yt,Aubert:2003bq,Cronin-Hennessy:kg}:
${\cal B}(B^+ \to \eta K^+) = (3.7 \pm 0.7) \times 10^{-6}$ and
${\cal B}(B^0 \to \eta K^0) = (2.9 \pm 1.0) \times 10^{-6}$~.

However, we note that the inputs given in Eq. (\ref{goodfitinputs})
provide large effects of $X_A$ and $X_H$: $e.g.$, $\rho^{PP}_{A,H}
= \rho^{VP}_{A,H} =1$ [see Eq. (\ref{XAXH})].
In order to explicitly estimate the effects of $X_A$ and $X_H$, we
also examine three interesting cases.  In the fourth column of
Table II, the BRs for $B \to \eta' K$ and $B \to \phi K$ are calculated
for $X_A =X_H =0$. Similarly, in the fifth and last column, those BRs
are calculated under the assumption of $X_A =0$ or $X_H =0$,
respectively.
We see that the contributions from the terms involving $X_A$ and
$X_H$ are quite large for both $B \to \eta' K$ and $B \to \phi K$
decays. In particular, for ${\cal B}(B^{+(0)} \to \phi K^{+(0)})$
the contribution of $X_A$ ($i.e.,$ weak annihilation contribution)
dominates over all the other contributions.
>From the table it is clear that the internal interference between
the effects of $X_A$ and $X_H$ on $B \to \eta' K$ is constructive,
while that on $B \to \phi K$ is destructive.
It should be stressed that in this scenario ($i.e.$, with the large effect
of $X_{A,H}$ allowed by the ``good'' fit), there is no room for invoking
new physics effects on the quark-level process $b\to s \bar s s$, which
is implied by the \emph{large negative} value of $\sin(2 \phi_1)$
recently measured by Belle \cite{Abe:2003yt}.

{\bf [Case 2] with the small $X_A$ and $X_H$ effects}

As already emphasized, if the nonpertubative contributions of $X_A$ and
$X_H$ are too large, the predictions based on these contributions become
less reliable and suspicious.  However, in {\bf [Case 1]}, we noticed that
the contribution of $X_A$ is very large, especially for
$B^{+(0)} \to \phi K^{+(0)}$ modes.
Therefore, it is natural to investigate presumably more reliable scenarios,
where the effects of $X_A$ and $X_H$ are rather small or at least not
dominant.

Using the global analysis for the twelve decay modes shown in Table I,
we find such a fit (with $\chi^2_{min} =18.3$) with the (relatively) small
$X_A$ and $X_H$ effects.
The corresponding theoretical inputs for this fit are as follows:
\begin{eqnarray}
&\mbox{}& \lambda =0.2198, ~~  A=0.868, ~~ \phi_3 =86.8^0, ~~
  |V_{ub}|=3.35 \times 10^{-3},
\nonumber \\
&\mbox{}& \mu =2.1 ~{\rm GeV}, ~~ m_s (2 ~{\rm GeV}) =85 ~{\rm MeV},
 ~~ f_B =220 ~{\rm MeV},
\nonumber \\
&\mbox{}& F^{B \pi} =0.249, ~~  R_{\pi K} =1, ~~ A^{B \rho} =0.31,
\nonumber \\
&\mbox{}&  \rho^{PP}_A =0, ~~ \rho^{VP}_A =0.5,
 ~~ \rho^{PP}_H =1, ~~ \rho^{VP}_H =0.746,
\nonumber \\
&\mbox{}& \phi^{VP}_A =-6^0, ~~ \phi^{VP}_H = \phi^{PP}_H =180^0.
\label{2ndfitinputs}
\end{eqnarray}
Note that in this case the effect of the weak annihilation parameter
$X_A$ is relatively small ($i.e.$, $\rho^{PP}_A =0$ and $\rho^{VP}_A =0.5$),
and the effect of the hard spectator scattering parameter $X_H$ is very
small, because $\rho^{PP}_H =1$, $\rho^{VP}_H =0.746$, and
$\phi^{PP}_H = \phi^{VP}_H =180^0$ so that the terms 1 and
$\rho_H e^{i \phi_H}$ in $X_H$ [see Eq. (\ref{XAXH})] cancel each other.

Based on the above inputs, the BRs for $B \to \eta' K$ and $B \to \phi K$
are predicted as
\begin{eqnarray}
&\mbox{}& {\cal B}(B^+ \to \eta' K^+) = 51.1 \times 10^{-6}, ~~~~
{\cal B}(B^0 \to \eta' K^0) = 46.8 \times 10^{-6},  \nonumber \\
&\mbox{}&  {\cal B}(B^+ \to \phi K^+) = 7.29 \times 10^{-6}, ~~~~
{\cal B}(B^0 \to \phi K^0) = 6.65 \times 10^{-6}.
\label{BRs}
\end{eqnarray}
These BRs are quite small, especially for $B \to \eta' K$, compared with
the experimental data, because of the small effects of $X_A$ and $X_H$
as well as the other fitted parameters such as $\phi_3$.
Since both processes $B \to \eta' K$ and $B \to \phi K$ have the same
(dominant) internal quark-level process $b \to s \bar s s$, we take into
account the possibility that there could be new physics effects on
the process $b \to s \bar s s$: for instance, as considered in
\cite{dko2,dkoz} in order to explain the large negative value of
$\sin(2 \phi_1)_{\phi K_s}$ reported by Belle.
We investigate whether it is possible to understand the difference
between the BRs given in Eq. (\ref{BRs}) and the experimental data,
by invoking new physics.

As specific examples, we consider two new physics scenarios: R-parity
violating (RPV) SUSY and R-parity conserving (RPC) SUSY.

{\bf (a) R-parity violating SUSY case}

The RPV part of the superpotential of the minimal supersymmetric standard
model can contain terms of the form
\begin{eqnarray}
 {\cal W}_{\rm RPV} &=& \kappa_iL_iH_2 + \l_{ijk}L_iL_jE_k^c
   + \l'_{ijk}L_iQ_jD_k^c + \l''_{ijk}U_i^cD_j^cD_k^c ~,
\label{superpot}
\end{eqnarray}
where $E_i$, $U_i$ and $D_i$ are respectively the $i$-th type
of lepton, up-quark and down-quark singlet superfields, $L_i$ and $Q_i$ are
the SU$(2)_L$ doublet lepton and quark superfields, and $H_2$ is the Higgs
doublet with the appropriate hypercharge.

For our purpose, we will assume only  $\l'-$type  couplings to be present.
Then, the effective Hamiltonian for charmless hadronic $B$ decay can be
written as \cite{Choudhury:1998wc},
\begin{eqnarray}
{H_{eff}^{\lambda'}} (b\ra \bar d_j d_k d_n)
   &=& d^R_{jkn} \left[ \bar d_{n\alpha} \gamma^\mu_L d_{j\beta}
          ~~ \bar d_{k\beta} \gamma_{\mu R} b_{\alpha} \right]
       + d^L_{jkn} \left[ \bar d_{n\alpha} \gamma^\mu_L b_{\beta}
          ~~ \bar d_{k\beta} \gamma_{\mu R} d_{j\alpha} \right],
\nonumber \\
{H_{eff}^{\lambda'}}(b\ra \bar u_j u_k d_n)
   &=& u^R_{jkn} \left[ \bar u_{k\alpha} \gamma^\mu_L u_{j\beta}
          ~~ \bar d_{n\beta} \gamma_{\mu R} b_{\alpha} \right].
\end{eqnarray}
Here the coefficients $d^{L,R}_{jkn}$ and $u^R_{jkn}$ are defined as
\begin{eqnarray}
d^R_{jkn} &=& \sum_{i=1}^3 {\l'_{ijk}\l'^{\ast}_{in3} \over 8\msnus}, ~~~
d^L_{jkn} =  \sum_{i=1}^3 {\l'_{i3k}\l'^{\ast}_{inj} \over 8\msnus}, ~~
(j,k,n=1,2)  \nonumber \\
u^R_{jkn} &=& \sum_{i=1}^3 {\l'_{ijn}\l'^{\ast}_{ik3} \over 8\msells}, ~~~
(j,k=1, \ n=2)
\end{eqnarray}
where $\alpha$ and $\beta$ are color indices and
$\gamma^\mu_{R, L} \equiv \gamma^\mu (1 \pm \gamma_5)$.
The leading order QCD correction to this operator is given by a scaling
factor $f\simeq 2$ for $m_{\tilde\nu}=200$ GeV.
We refer to Refs. \cite{Choudhury:1998wc,dko1} for the relevant notations.

The RPV SUSY part (relevant to the quark-level process $b \to s \bar s s$)
of the decay amplitude of $B^- \to \eta' K^-$ is given by
\begin{eqnarray}
{\cal A}^{\rm RPV}_{\eta' K} = {\left( d^L_{222} - d^R_{222} \right)}
 \left[  \frac{\bar m}{m_s} \left( A_{\eta'}^s -A_{\eta'}^u \right)
 \left( \tilde a_6 + {f^u_{\eta'} \over f^s_{\eta'}} \tilde a'_6 \right)
 + A_{\eta'}^s ( \tilde a_4 -\tilde a_5 ) + A_{\eta'}^u \tilde a_4 \right],
\end{eqnarray}
where
\begin{eqnarray}
\bar m \equiv {m^2_{\eta'} \over (m_b-m_s)}, ~~~
A^{u(s)}_{\eta'} = f^{u(s)}_{\eta'} F^{B \to K} (m_B^2 -m_K^2).
\end{eqnarray}
Here the coefficients $\tilde a^{(\prime)}_i$ are expressed as
\begin{eqnarray}
\tilde a_4 &=& {C_F \alpha_s \over 4 \pi N_c}
 \left[ {4  \over 3} \ln{m_b \over \mu} - G_K(0) \right],
 \nonumber \\
\tilde a_5 &=& {1 \over N_c} \left[ 1 - {C_F \alpha_s \over 4 \pi}
 \left( V_{\eta'} +12 +{4 \pi^2 \over N_c} H(BK, \eta') \right) \right],
 \nonumber \\
\tilde a_6 &=& 1 + {C_F \alpha_s \over 4 \pi N_c}
 \left[ {4  \over 3} \ln{m_b \over \mu} - \hat G_K(0) \right],
 \nonumber \\
\tilde a^{\prime}_6 &=& {C_F \alpha_s \over 4 \pi N_c}
 \left[ {4  \over 3} \ln{m_b \over \mu} - \hat G_K(0) \right],
\end{eqnarray}
where $G_K(0) ={5 \over 3} +{2 \pi \over 3}i$ and
$\hat G_K(0) ={16 \over 9} +{2 \pi \over 3}i$.

It has been noticed \cite{dko2} that ${\cal A}^{\rm RPV}_{\eta' K}$ is
proportional to $( d^L_{222} -d^R_{222})$, while the RPV part of
the decay amplitude of $B \to \phi K$ is proportional to
$( d^L_{222} +d^R_{222})$.
It has been also pointed out \cite{dko2} that the opposite relative
sign between $d^L_{222}$ and $d^R_{222}$ in the modes $B \to \eta' K$
and $B \to \phi K$ appears due to the different parity in the final
state mesons $\eta'$ and $\phi$, and this different combination of
$( d^L_{222} -d^R_{222})$ and $( d^L_{222} +d^R_{222})$ in these
modes plays an important role to explain both the large BRs for
$B \to \eta' K$ and the large negative value of
$\sin(2 \phi_1)_{\phi K_s}$ at the same time.

We define the new coupling terms $d^L_{222}$ and $d^R_{222}$ as follows:
\begin{eqnarray}
d^L_{222} \propto | \lambda^{\prime}_{i32}
\lambda^{\prime *}_{i22}| e^{i \theta_L}~, ~~
d^R_{222} \propto |\lambda^{\prime}_{i22} \lambda^{\prime *}_{i23}|
e^{i \theta_R},
\end{eqnarray}
where $\theta_L$ and $\theta_R$ denote new weak phases of the product
of new couplings
$\lambda^{\prime}_{i32} \lambda^{\prime *}_{i22}$ and
$\lambda^{\prime}_{i22} \lambda^{\prime *}_{i23}$, respectively, as defined
by $\lambda^{\prime}_{332} \lambda^{\prime *}_{322} \equiv
| \lambda^{\prime}_{332} \lambda^{\prime *}_{322}| e^{i \theta_L}$ and
$\lambda^{\prime}_{322} \lambda^{\prime *}_{323} \equiv
|\lambda^{\prime}_{322} \lambda^{\prime *}_{323}| e^{i \theta_R}$.
We find that the experimental measurements of the BRs for
$B^{+(0)} \to \eta' K^{+(0)}$ and $B^{+(0)} \to \phi K^{+(0)}$ can be
consistently understood for the following values of the parameters:
\begin{eqnarray}
 |\lambda^{\prime}_{322}| = 0.076 ~,~ |\lambda^{\prime}_{332}| = 0.076 ~,~
 |\lambda^{\prime}_{323}| = 0.064~, \nonumber \\
 \theta_L = 1.32 ~,~ \theta_R = -1.29~,~~ m_{\rm SUSY} = 200 ~{\rm GeV}.
\label{rpv}
\end{eqnarray}
Our results are summarized in Table III.
In addition to the parameters given in Eq. (\ref{rpv}), we also used the
additional strong phase $\delta' =30^0$, which can arise from the power
contributions of $\Lambda_{QCD}/ m_b$ neglected in the QCDF scheme,
and whose size can be in principle comparable to the strong phase arising
from the radiative corrections of $O(\alpha_s)$.
It has been shown \cite{dkoz} that using $\delta' =30^0$ together with
the parameters given in Eq. (\ref{rpv}), one can explain the large
negative value of $\sin(2 \phi_1)_{\phi K_s}$ as well.
Notice that the new coupling terms $d^L_{222}$ and $d^R_{222}$ are
relevant only to the process $b \to s \bar s s$, so they do not affect
other $B \to PP$ and $VP$ decays, such as $B \to \pi \pi$, $\pi K$,
$\rho \pi$, $\rho K$, etc, which are already well understood within
the SM.

In the case of the large $X_{A,H}$ effects with $\chi^2_{\min}=7.5$
where the BR of $B^+ \to \eta' K^+$ is large, we can use the R-parity
violating SUSY couplings to raise the BR of $B^+ \to \phi K^+$ (which
is small, $4.02 \times 10^{-6}$ to begin with).
It is possible to raise ${\cal B}(B^+ \to \phi K^+)$ to $(8-9) \times
10^{-6}$.  However, in this case, $\sin{(2\phi_1)}_{\phi K_s}$ can not
be large negative \cite{dkoz}.

The RPV terms can arise in the context of SO(10) models which explain the
small neutrino mass and has an intermediate breaking scale where $B-L$
symmetry gets broken by $(16+\bar{16})$ Higgs. These additional Higgs form
operators like $16_H16_m16_m16_m/M_{pl}$ (16$_m$ contains matter fields)
and generate the RPV terms \cite{Mohapatra:1996pu}.

\begin{table}
\caption{The branching ratios (in unit of $10^{-6}$) for $B \to \eta' K$
and $B \to \phi K$ decays calculated in the framework of R-parity
violating SUSY.  Here the inputs for the fit with small $X_A$ and $X_H$
are used. }
\smallskip
\begin{tabular}{c|c||c|c}
\hline \hline
Decay mode & Prediction  & Decay mode & Prediction
\\ \hline
   $B^+ \to \eta' K^+ $ & 74.0 & $B^0 \to \eta' K^0 $ & 67.7
\\ $B^+ \to \phi K^+$   & 10.2 & $B^0 \to \phi K^0$   & 9.5
\\ \hline \hline
\end{tabular}
\end{table}

{\bf (b) R-parity conserving SUSY case}

As an example of the RPC SUSY case, we will consider the supergravity
(SUGRA) model with the simplest possible non-universal soft terms which
is the simplest extension of the minimal SUGRA (mSUGRA) model. In this model the
lightest SUSY particle is stable and this particle can explain the
dark matter content of the universe. The recent WMAP result provides\cite{wmap}:
\begin{equation}\Omega_{\rm CDM}h^2=0.1126^{+0.008}_{-0.009},\end{equation} and
we implement 2$\sigma$ bound in our calculation.

In the SUGRA model, the superpotential and soft SUSY breaking terms
at the grand unified theory (GUT) scale are given by
\begin{eqnarray}
{\cal W} &=& Y^U Q H_2 U + Y^D Q H_1 D + Y^L L H_1 E + \mu H_1 H_2 ,
\nonumber \\
{\cal L}_{\rm soft} &=& - \sum_i m_i^2 |\phi_i|^2
 - \left[ {1 \over 2} \sum_{\alpha} m_{\alpha} \bar \lambda_{\alpha}
 \lambda_{\alpha} + B \mu H_1 H_2 \right.  \nonumber \\
&\mbox{}& \left. + (A^U Q H_2 U + A^D Q H_1 D + A^L L H_1 E)
  + {\rm H.c.} \right],
\end{eqnarray}
where $E$, $U$ and $D$ are respectively the lepton, up-quark and
down-quark singlet superfields, $L$ and $Q$ are the SU$(2)_L$ doublet
lepton and quark superfields, and $H_{1,2}$ are the Higgs doublets.
$\phi_i$ and $\lambda_{\alpha}$ denote all the scalar fields and gaugino
fields, respectively.
In the mSUGRA model, a universal scalar mass $m_0$, a universal gaugino
mass $m_{1/2}$, and the universal trilinear coupling $A$ terms are
introduced at the GUT scale:
\begin{equation}
m_i^2 = m_0^2, ~~~ m_{\alpha} = m_{1/2}, ~~~
A^{U,D,L} = A_0 Y^{U,D,L},
\end{equation}
where $Y^{U,D,L}$ are the diagonalized $3 \times 3$ Yukawa matrices.
In this model, there are four free parameters, $m_0$, $m_{1/2}$, $A_0$,
and $\tan \beta \equiv \langle H_2 \rangle / \langle H_1 \rangle$, in
addition to the sign of $\mu$.  The parameters $m_{1/2}$, $\mu$ and
$A$ can be complex, and four phases appear: $\theta_A$ (from $A_0$),
$\theta_1$ (from the gaugino mass $m_1$), $\theta_3$ (from the gaugino
mass $m_3$), and $\theta_{\mu}$ (from the $\mu$ term).

It has been shown in Refs. \cite{Arnowitt:2003ev,Khalil:2002fm}
that the mSUGRA model can not explain the large negative value of
$\sin(2 \phi_1)_{\phi K_s}$, because in this model the only source
of flavor violation is in the CKM matrix, which can not provide a
sufficient amount of flavor violation needed for the $b \to s$
transition in the processes $B \to \phi K$.
The minimal extension of the mSUGRA has been studied to solve the
large negative $\sin(2\phi_1)_{\phi K_s}$ in the context of QCDF
\cite{Arnowitt:2003ev}, or both large negative
$\sin(2\phi_1)_{\phi K_s}$ and large BR of $B \to \eta' K$ in the
context of NF \cite{Khalil:2002fm}.

The minimal extension of the mSUGRA model contains non-universal soft
breaking $A$ terms, in addition to the parameters in the mSUGRA model.
In order to enhance contributions to the $b \to s$ transition, the
simplest choice is to consider only non-zero (2,3) elements in $A$ terms
which enhance the left-right mixing of the second and third generation.
The $A$ terms with only non-zero (2,3) elements can be expressed as
\begin{equation}
A^{U,D} = A_0 Y^{U,D} + \Delta A^{U,D},
\end{equation}
where $\Delta A^{U,D}$ are $3 \times 3$ complex matrices and
$\Delta A^{U,D}_{ij} = \left| \Delta A^{U,D}_{ij} \right|
e^{i\phi^{U,D}_{ij}}$ with $\left| \Delta A^{U,D}_{ij} \right| = 0$
unless $(i,j)= (2,3)~{\rm or}~(3,2)$.
It is obvious that the mSUGRA model is recovered if $\Delta A^{U,D}=0$.

For our analysis, we consider all the known experimental constraints
on the parameter space of the model, as in Ref. \cite{Arnowitt:2003ev}.
Those constraints come from the radiative $B$ decay process
$B \to X_s \gamma$ ($2.2 \times 10^{-4} < {\cal B}(B \to X_s \gamma)
<4.5\times 10^{-4}$ \cite{Alam,bsg}),
neutron and electron electric dipole moments
($d_{n} < 6.3 \times 10^{-26} e ~cm$, $d_{e} < 0.21 \times 10^{-26}
 e ~cm$ \cite{pdg}), relic density measurements, $K^0 - \bar K^0$ mixing
($\Delta M_K = (3.490 \pm 0.006) \times 10^{-12}$ MeV \cite{pdg}),
LEP bounds on masses of SUSY particles and the lightest Higgs
($m_h\geq 114$ GeV).
From the experimental constraints, we find that $\theta_1 \approx
22^0$, $\theta_3 \approx 30^0$, and $\theta_{\mu} \approx -11^0$.
For the phase $\theta_A$, we set $\theta_A = \pi$.
It has been noticed \cite{Arnowitt:2003ev} that the SUSY contribution
mainly affects the Wilson coefficients $C_{8g(7\gamma)}$ and
$\tilde C_{8g(7\gamma)}$ and these coefficients do not change the weak
annihilation effects arising from the SM calculation.

\begin{table}
\caption{The branching ratios (in units of $10^{-6}$)
for $B^+ \to \eta' K^+$ (left) and $B^+ \to \phi K^+$ (right)
at $\tan\beta=10$ with non-zero $\Delta A^D_{23}$ and
$\Delta A^D_{32}$.  The units for $m_{1/2}$, $|A_0|$, and
$\left| \Delta A^D_{23(32)} \right|$ are in GeV. }
\smallskip
\begin{tabular}{c|c|c|c|c|c|c|c|c|c}
\hline \hline
$|A_0|$ & \multicolumn{2}{c|}{$800$} & \multicolumn{2}{c|}{$600$}
& \multicolumn{2}{c|}{$400$} & \multicolumn{2}{c|}{$0$}
& $\left| \Delta A^D_{23(32)} \right|$
\\ \hline
$m_{1/2}=300$ & $\ba{c} 79.6 \ea$ & $\ba{c} 9.9 \ea$
 & $\ba{c} 81.0 \ea$ & $\ba{c} 9.2 \ea$
 & $\ba{c} 79.6 \ea$ & $\ba{c} 9.1 \ea$
 & $\ba{c} 79.0 \ea$ & $\ba{c} 8.1 \ea$
 & $66 - 74$
\\ \hline
$m_{1/2}=400$ & $\ba{c} 78.2 \ea$ & $\ba{c} 9.9 \ea$
 & $\ba{c} 83.0 \ea$ & $\ba{c} 9.6 \ea$
 & $\ba{c} 79.0 \ea$ & $\ba{c} 9.2 \ea$
 & $\ba{c} 81.0 \ea$ & $\ba{c} 8.5 \ea$
 & $150 - 168$
\\ \hline
$m_{1/2}=500$ & $\ba{c} 84.8 \ea$ & $\ba{c} 9.9 \ea$
 & $\ba{c} 83.7 \ea$ & $\ba{c} 9.9 \ea$
 & $\ba{c} 81.0 \ea$ & $\ba{c} 10.0 \ea$
 & $\ba{c} 77.0 \ea$ & $\ba{c} 8.1 \ea$
 & $244 - 256$
\\ \hline
$m_{1/2}=600$ & $\ba{c} 73.0 \ea$ & $\ba{c} 7.6 \ea$
 & $\ba{c} 71.0 \ea$ & $\ba{c} 7.5 \ea$
 & $\ba{c} 70.0 \ea$ & $\ba{c} 7.5 \ea$
 & $\ba{c} 70.0 \ea$ & $\ba{c} 7.1 \ea$
 & $270 - 304$
\\ \hline \hline
\end{tabular}
\end{table}

In our calculation, we consider the case with non-zero $\Delta A^D_{23}$
and non-zero $\Delta A^D_{32}$ for $\tan \beta =10$.  All the other
elements in $\Delta A^{U,D}$ are set to be zero. We
compute the BRs for $B \to \eta' K$ and $B \to \phi K$ in the case of
$\left| \Delta A^D_{23} \right| \sim\left| \Delta A^D_{32} \right|$
and $\phi^D_{23} \neq \phi^D_{32}$ with $\tan \beta =10$.
Table IV shows the BRs for $B^+ \to \eta' K^+$ and $B^+ \to \phi K^+$
calculated for various values of the parameters $m_{1/2}$, $|A_0|$
and $|\Delta A^D_{23(32)}|$.
For each $m_{1/2}$ and $|A_0|$, the left column shows the BR for
$B^+ \to \eta' K^+$ and the right column shows the BR for
$B^+ \to \phi K^+$.  All the predicted BRs in the table are well
consistent with the experimental data.
The BR for $B^+ \to \eta K^+$ is estimated as $(3.1 \sim 4.4)
\times 10^{-6}$ which also agrees with the data.
The higher $\tan\beta$ values are also allowed, but the allowed range
of $m_{1/2}$ becomes smaller.  We satisfy the relic density constraint
using the stau--neutralino co-annihilation channel \cite{cnst}.

For the numerical calculation, we used the QCD parameters given in
Eq. (\ref{2ndfitinputs}) and the additional strong phase $\delta' =0$.
The value of $m_{1/2}$ varies from 300 GeV to 600 GeV, and the value
of $|A_0|$ varies from 0 to 800 GeV.  Even though the value of $m_0$
is not explicitly shown, it is chosen for different $m_{1/2}$ and $A_0$
such that the relic density constraint is satisfied, $e.g.,$ for
$m_{1/2}$=300 GeV, $m_0$ varies in the range ($70-110$) GeV.  The value of
$m_0$ increases as $m_{1/2}$ increases.
The value of $\left| \Delta A^D_{23(32)} \right|$ increases as
$m_{1/2}$ does.  The phases $\phi^D_{23}$ and $\phi^D_{32}$ are
approximately $-40^0$ to $-15^0$ and $165^0$ to $180^0$, respectively.
So far we have assumed that $\Delta A^U_{23,32}=0$. But if we use $\Delta
A^U_{23,32}\neq 0$ and $\Delta A^D_{23,32}=0$, the value of
$\sin(2\phi)_{\phi K_s}$ is mostly positive.

In passing, we note that the set of the same parameters used in our
calculation can also produce the large negative value of
$\sin (2 \phi_1)_{\phi K_s}$ \cite{dkoz}. As a final comment, we note
that in the case of the large $X_{A,H}$ effects with
$\chi^2_{\min}=7.5$, it is possible to raise the BR for $B^+ \to
\phi K^+$ to $(8 - 9) \times 10^{-6}$. However, in that case, the
large negative value of $\sin(2\phi_1)_{\phi K_s}$ can not be
obtained \cite{dkoz}.

\section{Conclusion}

We investigated the decay processes $B^{+(0)} \to \eta' K^{+(0)}$ in
the QCDF approach.
In order to reliably estimate the weak annihilation parameter
$X_A \equiv \left( 1 + \rho_A e^{i \phi_A} \right) {\rm ln}
{m_B \over \Lambda_h}$
and the hard spectator scattering parameter $X_H \equiv \left( 1 +
\rho_H e^{i \phi_H} \right) {\rm ln} {m_B \over \Lambda_h}$
arising from logarithmic
divergences in the end-point region, we used the global analysis
for twelve $B \to PP$ and $VP$ decay modes, such as $B \to \pi \pi$,
$\pi K$, $\rho \pi$, $\rho K$, $\omega \pi$, $\omega K$.
{}From the global analysis, we found that both the large effect of $X_{A,H}$
(less reliable) and the small effect of $X_{A,H}$ (more reliable) are
allowed.
For the former case, the parameters $\rho_{A,H}$ and $\phi_{A,H}$ are
determined to be: $\rho^{PP}_A = \rho^{VP}_A = \rho^{PP}_H =
\rho^{VP}_H =1$, $\phi^{PP}_A =57^0,~ \phi^{VP}_A =52^0,~
\phi^{PP}_H =-23^0,~ \phi^{VP}_H =180^0$.  For the latter case, the
parameters $\rho_{A,H}$ and $\phi_{A,H}$ are: $\rho^{PP}_A =0,~
\rho^{VP}_A =0.5,~ \rho^{PP}_H =1,~ \rho^{VP}_H =0.746$,
$\phi^{VP}_A =-6^0,~ \phi^{VP}_H = \phi^{PP}_H =180^0$.

In the case of the large $X_{H,A}$ effects allowed by the ``good'' fit
(with $\chi^2_{min} =8.6$ for the twelve decay modes), the BRs for
$B^{+(0)} \to \eta' K^{+(0)}$ and $B^{+(0)} \to \phi K^{+(0)}$ calculated
within the SM saturate the large values of the experimental results
measured by Belle, BaBar, and CLEO.  Thus, there is no room for invoking
new physics effects on the quark-level process $b \to s \bar s s$, which
are implied by the large negative value of $\sin (2 \phi_1)_{\phi K_s}$
recently reported by Belle.

In contrast, in the case of the small $X_{H,A}$ effects that is
theoretically more reliable, the SM prediction for these BRs is smaller
than the experimental data.
Since both $B^{+(0)} \to \eta' K^{+(0)}$ and $B^{+(0)} \to \phi K^{+(0)}$
have the same (dominant) internal process $b \to s \bar s s$, we took into
account possible new physics effects on the $b \to s \bar s s$ transition,
as in \cite{dko2,dkoz} for explaining the recent Belle measurement of
$\sin (2\phi_1)_{\phi K_s}$.  Specifically, we considered two new physics
scenarios: R-parity violating SUSY and R-parity conserving SUSY.
In the RPV SUSY case, the BRs for $B^{+(0)} \to \eta' K^{+(0)}$ are
predicted as $73.9 (67.8) \times 10^{-6}$ and the BRs for
$B^{+(0)} \to \phi K^{+(0)}$ are $10.2 (9.5) \times 10^{-6}$ which are
consistent with the data.
The relevant new couplings are found to be:
$ |\lambda^{\prime}_{322}| = 0.086,~ |\lambda^{\prime}_{332}| = 0.089,~
|\lambda^{\prime}_{323}| = 0.030,~ \theta_L = 0.66,~ \theta_R = -2.25$.
As an example of the RPC SUSY case, we adopted the simplest extension
of the mSUGRA model, which contains only non-zero (2,3) elements in the
soft breaking trilinear coupling $A$ terms, in addition to the other
parameters of the mSUGRA model.  Considering all the known constraints
on the relevant parameter space, we found that for $\tan \beta =10$,
${\cal B}(B^+ \to \eta' K^+) = (70.0 \sim 84.8) \times 10^{-6}$ and
${\cal B}(B^+ \to \phi K^+) = (7.1 \sim 10.0) \times 10^{-6}$, which
are in good agreement with the experimental data.

\vspace{1cm}
\centerline{\bf ACKNOWLEDGEMENTS}

\noindent The work of B.D was supported by Natural Sciences and
Engineering Research Council of Canada. The work of C.S.K. was
supported in part by Grant No. R02-2003-000-10050-0 from BRP of
the KOSEF and in part by CHEP-SRC Program. The work of S.O. and
G.Z. was supported by the Japan Society for the Promotion of
Science (JSPS).
\\


\end{document}